\documentclass[a4paper, fleqn, usenatbib, useAMS]{mnras}

\usepackage{graphicx} 
\usepackage{amsmath, amssymb} 
\usepackage{newtxtext, newtxmath} 
\usepackage[T1]{fontenc}
\usepackage{ae, aecompl}
\usepackage{xcolor}

\title[Planet-induced radio emission from M dwarfs]{Planet-induced radio emission from the coronae of M dwarfs: the case of Prox Cen and AU Mic}

\author[Kavanagh et al.]{Robert D. Kavanagh$^{1}$\thanks{Contact e-mail: \href{mailto:kavanar5@tcd.ie}{kavanar5@tcd.ie}}, Aline A. Vidotto$^{1}$, Baptiste Klein$^{2}$, Moira M. Jardine$^{3}$,\newauthor Jean-Fran\c{c}ois Donati$^{4}$, D\'ualta \'O Fionnag\'ain$^{5}$ \\ 
$^{1}$School of Physics, Trinity College Dublin, The University of Dublin, Dublin 2, Ireland \\
$^{2}$Sub-department of Astrophysics, Department of Physics, University of Oxford, Oxford OX1 3RH, UK \\
$^{3}$SUPA, School of Physics and Astronomy, University of St Andrews, St Andrews KY16 9SS, UK \\
$^{4}$Universit\'e de Toulouse, CNRS, IRAP, 14 av. Belin, 31400 Toulouse, France \\
$^{5}$Centre for Astronomy, National University of Ireland, Galway, Ireland}

\date{Last updated ...; in original form ...}

\pubyear{2020}

\graphicspath{{figs/}}

\begin{document}
\label{firstpage}
\pagerange{\pageref{firstpage}--\pageref{lastpage}}
\maketitle


\begin{abstract}
There have recently been detections of radio emission from low-mass stars, some of which are indicative of star-planet interactions. Motivated by these exciting new results, in this paper we present Alfv\'en wave-driven stellar wind models of the two active planet-hosting M dwarfs Prox Cen and AU Mic. Our models incorporate large-scale photospheric magnetic field maps reconstructed using the Zeeman-Doppler Imaging method. We obtain a mass-loss rate of $0.25~\dot{M}_{\sun}$ for the wind of Prox Cen. For the young dwarf AU Mic, we explore two cases: a low and high mass-loss rate. Depending on the properties of the Alfv\'en waves which heat the corona in our wind models, we obtain mass-loss rates of $27$ and $590~\dot{M}_{\sun}$ for AU Mic. We use our stellar wind models to assess the generation of electron cyclotron maser instability emission in both systems, through a mechanism analogous to the sub-Alfvénic Jupiter-Io interaction. For Prox Cen we do not find any feasible scenario where the planet can induce radio emission in the star's corona, as the planet orbits too far from the star in the super-Alfv\'enic regime. However, in the case that AU Mic has a stellar wind mass-loss rate of $27~\dot{M}_{\sun}$, we find that both planets b and c in the system can induce radio emission from $\sim10$~MHz -- 3~GHz in the corona of the host star for the majority of their orbits, with peak flux densities of $\sim10$~mJy. Detection of such radio emission would allow us to place an upper limit on the mass-loss rate of the star.
\end{abstract}



\section{Introduction}

Many theoretical works have aimed to identify potential targets for the detection of exoplanetary radio emission \citep{griessmeier07, vidotto15, vidotto17, kavanagh19}. A model that is commonly used to estimate the generation of radio emission from exoplanets is the radiometric Bode's law, which extrapolates the observed relation between the radio power emitted from the magnetised solar system objects and the incident solar wind power \citep{zarka07}. In this model, the interaction between the planet's magnetosphere and the magnetised stellar wind of its host star leads to reconnection in the planet's magnetotail. This reconnection then accelerates electrons towards the planet's poles, creating emission through the electron cyclotron maser instability (ECMI). 

In these works, hot Jupiters have often been selected as suitable candidates for radio detection. As they are exposed to harsh stellar wind conditions due to their proximity to their host stars \citep{vidotto15}, they are expected to produce much higher radio powers than those observed in the solar system \citep{zarka01}. This emission can only be generated if hot Jupiters are magnetised. Given that the emission mechanism is ECMI, it occurs at the cyclotron frequency $\nu = 2.8~B$~MHz, where $B$ is the magnetic field strength in gauss (G). \citet{zaghoo18} estimated that hot Jupiters could harbour magnetic fields with strengths of 1 -- 10~G. In this range, these field strengths would correspond to radio emission from 2.8 -- 28~MHz.

Despite hot Jupiters being seemingly favourable sources of detectable low frequency radio emission, many radio surveys of these systems have found little evidence of such emission \citep{smith09, lazio10, lecavelier13, degasperin20, narang21}. Numerous explanations have been proposed for the lack of detections, such as a mismatch between the emitted and observation frequency \citep{bastian00}, the emission being beamed out of our line of sight \citep{smith09}, and absorption of the emission by the stellar wind of the host star \citep{vidotto17, kavanagh19, kavanagh20}. Intriguingly, \citet{turner21} recently detected radio emission at $\sim30$~MHz from the hot Jupiter host $\tau$~Boo. They suggest that this emission could be indicative of ECMI generated in the orbiting planet's magnetic field. Follow-up observations will be required however in order to investigate this further.

An alternative unipolar model for the generation of radio emission via ECMI in exoplanetary systems has also been proposed, analogous to the sub-Alfv\'enic interaction between Jupiter and its moon Io \citep{neubauer80, zarka98, saur04, zarka07, griessmeier07}. In this model, the host star and planet take the roles of Jupiter and Io respectively. If the planet orbits with a sub-Alfv\'enic velocity relative to the wind of its host star, it can generate Alfv\'en waves that travel back towards the star \citep{ip04, mcivor06, lanza12, turnpenney18, strugarek19, vedantham20}. A fraction of the wave energy produced in this interaction is expected to dissipate and produce radio emission via ECMI in the corona of the host star \citep{turnpenney18}.

Due to the increasing sensitivity of radio telescopes such as LOFAR, M~dwarfs are beginning to light up the radio sky at low frequencies (Callingham et al., submitted). One such system is the quiescent M~dwarf GJ~1151, which was recently detected to be a source of 120 -- 160~MHz emission by \citet{vedantham20}. The authors illustrated that the observed emission is consistent with ECMI from the star induced by an Earth-sized planet orbiting in the sub-Alfv\'enic regime with a period of 1 -- 5 days. Prior to this detection, there had been no evidence to suggest GJ~1151 is host to a planet. There has been some discussion in the literature recently about the existence of such a planet. \citet{mahadevan21} have suggested that a planet orbits the star in a 2-day orbit, whereas \citet{perger21} have ruled this out, placing a mass upper limit of 1.2 Earth on a planet in a 5-day orbit. Follow-up observations of the system will be needed to further assess if the radio emission is of a planet-induced origin.

M~dwarfs typically exhibit strong magnetic fields, with maximum large-scale strengths ranging from $\sim100$~G -- 1~kG \citep{donati08, morin10, shulyak19}. If an M~dwarf is host to a planet that orbits in the sub-Alfv\'enic regime, the planet could induce radio emission from the star's corona via ECMI at frequencies up to 280~MHz -- 2.8~GHz. Planet-induced emission from M~dwarfs therefore could be distinguished from emission predicted with the radiometric Bode's law, due to the different frequency ranges. 

The M dwarf Proxima Centauri (Prox Cen) is our closest stellar neighbour, which also hosts a planet that is expected to be Earth-sized \citep{angladaescude16}. Naturally, being so close to us we want to understand both the habitability of the orbiting planet and the system's potential for producing planet-induced radio emission. An important step in answering this question is to assess the stellar wind environment around the planet. Prox Cen is expected to have a relatively weak stellar wind, with a mass-loss rate of $\dot{M}<0.2~\dot{M}_{\sun}$ \citep{wood01}, where $\dot{M}_{\sun} = 2\times10^{-14}~M_{\sun}~\textrm{yr}^{-1}$ is the solar wind mass-loss rate. The star also possesses a strong large-scale surface magnetic field of $\sim200$~G \citep{klein21}. Its proximity to Earth along with its potentially habitable planet make it an interesting system to search for planet-induced radio emission.

AU Microscopii (AU Mic) is a young M dwarf which also shows potential for detecting planet-induced radio emission. It lies just under 10~pc away from Earth, and has recently been discovered to host two Neptune-sized close-in planets \citep{plavchan20, martioli21}. We list the properties of the AU Mic system along with those for Prox Cen in Table~\ref{tab:m dwarf params}. While planets b and c orbiting AU Mic are not likely to be habitable, their proximity to the host star makes them ideal candidates for inducing radio emission in the corona of the host star. The stellar wind mass-loss rate of AU Mic is relatively unconstrained. Models of interactions between the stellar wind and debris disk in the system estimate a mass-loss rate from 10~$\dot{M}_{\sun}$ \citep{plavchan09} up to 1000~$\dot{M}_{\sun}$ \citep{chiang17}. Its large-scale surface magnetic field is also quite strong, with a strength of $\sim500$~G \citep{klein20}. The two close-in planets along with the star's strong magnetic field make it a very suitable candidate for planet-induced radio emission.

In this paper, we use the magnetic field maps reconstructed by \citet{klein21} and \citet{klein20} to model the stellar winds of the two M dwarfs Prox Cen and AU Mic respectively. We use our 3D Alfv\'en wave-driven stellar wind models to investigate if the orbiting planets can induce the generation of ECMI emission in the coronae of their host stars. We then illustrate how the detection of planet-induced radio emission can be used to constrain properties of the wind of the host star.

\begin{table*}
\caption{Stellar and planetary parameters for the Prox Cen and AU Mic systems.}
\label{tab:m dwarf params}
\centering
\begin{tabular}{lccccc}
\hline
Parameter & Prox Cen & Reference & \multicolumn{2}{c}{AU Mic} & Reference \\
\hline
Stellar mass ($M_\star$) & 0.12 $M_{\sun}$ & 1 & \multicolumn{2}{c}{0.50 $M_{\sun}$} & 2 \\
Stellar radius ($R_\star$) & 0.14 $R_{\sun}$ & 1 & \multicolumn{2}{c}{0.75 $R_{\sun}$} & 2 \\
Stellar wind mass-loss rate ($\dot{M}$) & < 0.2 $\dot{M}_{\sun}$ & 3 & \multicolumn{2}{c}{$10~\dot{M}_{\sun}$, $1000~\dot{M}_{\sun}$} & 4, 5 \\
Average surface stellar magnetic field strength & 200~G & 6 & \multicolumn{2}{c}{500~G} & 7 \\
Spectral type & M5.5V & 1 & \multicolumn{2}{c}{M1V} & 8 \\
Stellar rotation period & 83 days & 1 & \multicolumn{2}{c}{4.86 days} & 2 \\
Distance ($d$) & 1.30 pc & 1 & \multicolumn{2}{c}{9.79 pc} & 2 \\
\\
& \underline{Planet b} & & \underline{Planet b} & \underline{Planet c} \\
Planetary mass ($M_\textrm{p}$) & > 1.27 $M_\textrm{\earth}$ & 1 & 1.00 $M_\textrm{Nep}$ & 1.7 -- 27.7 $M_\textrm{\earth}$ & 9 \\
Planetary radius ($R_\textrm{p}$) & - & - & 1.13 $R_\textrm{Nep}$ & 0.91 $R_\textrm{Nep}$ & 9 \\
Planetary orbital distance ($a$) & 74.0 $R_\star$ & 1 & 18.5 $R_\star$ & 26.9 $R_\star$ & 9 \\
Planetary orbital period & 11.2 days & 1 & 8.46 days & 18.9 days & 9 \\
\hline
\multicolumn{6}{p{14cm}}{1: \cite{angladaescude16}; 2: \cite{plavchan20}; 3: \cite{wood01}; 4: \cite{plavchan09}; 5: \cite{chiang17}; 6: \cite{klein21}; 7; \cite{klein20}; 8: \cite{martioli20}; 9: \cite{martioli21}} \\
\hline
\end{tabular}
\end{table*}


\section{Modelling the winds of M-dwarfs}
\label{sec:modelling}

To model the stellar winds of Prox Cen and AU Mic we use the Alfv\'en wave solar model \citep[AWSoM,][]{vanderholst14} implemented in the 3D magnetohydrodynamic (MHD) code BATS-R-US \citep{powell99}, which is part of the Space Weather Modelling Framework \citep{toth12}. In the AWSoM model, Alfv\'en waves are injected at the base of the chromosphere where they propagate, reflect, and dissipate. The dissipation of the Alfv\'en waves heats the corona, which in turn drives the stellar wind outflow. The AWSoM model has been validated against observations of the solar corona \citep{vanderholst14, jin17}, and has also been applied to study the stellar wind environments of other systems \citep[see][]{garraffo17, borosaikia20, alvaradogomez20, ofionnagain21}.

\begin{figure}
\includegraphics[width = \columnwidth]{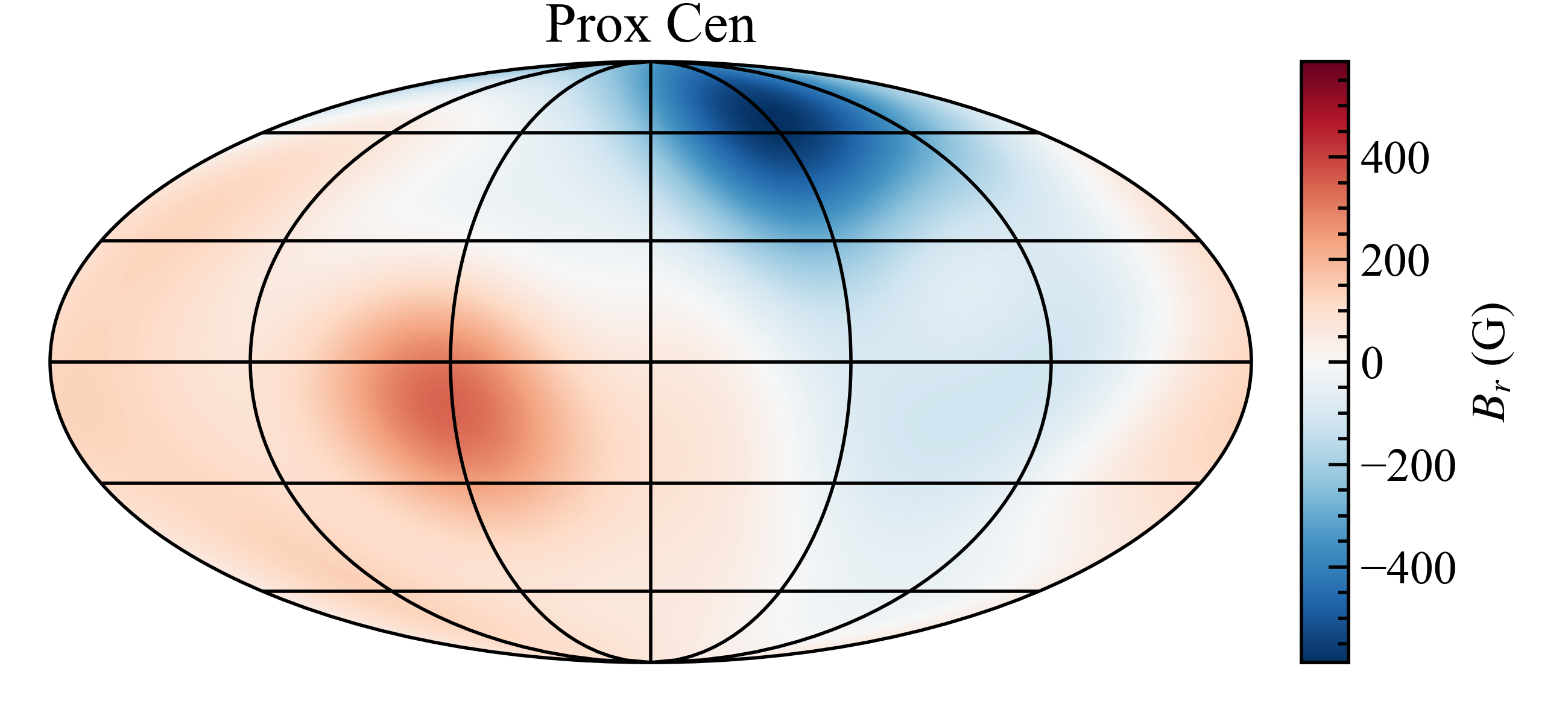} \\
\includegraphics[width = \columnwidth]{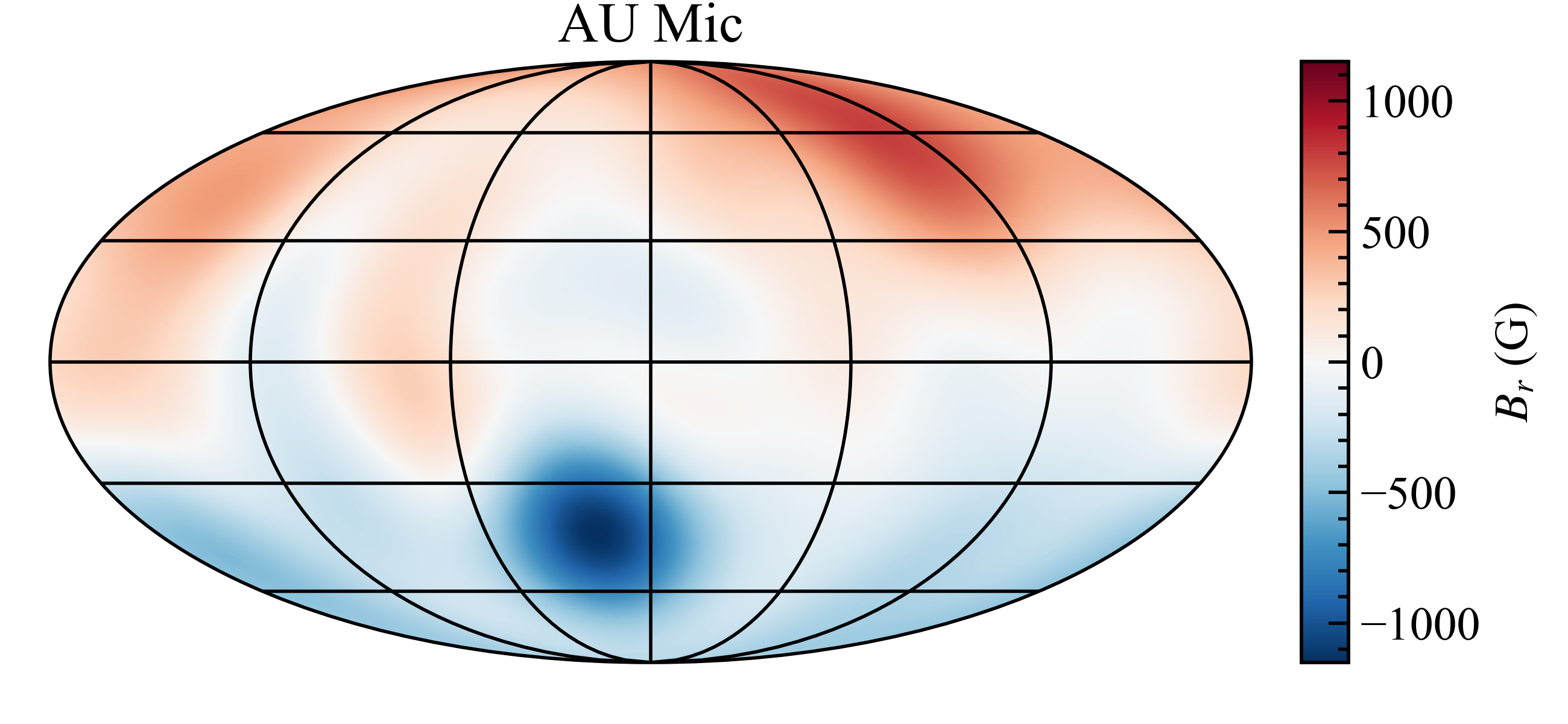}
\caption{Radial photospheric magnetic field of Prox Cen (top) and AU Mic (bottom), reconstructed by \citet{klein21} and \citet{klein20} respectively. We implement these maps at the inner boundary of our stellar wind simulations for the two stars.}
\label{fig:bfield maps}
\end{figure}

In our models, we set boundary conditions for the radial magnetic field, Alfv\'en wave flux, plasma density and temperature at the base of the chromosphere. For the magnetic field we use maximum-entropy maps reconstructed from spectropolarimetric observations for the two stars by \citet{klein21} and \citet{klein20} via the Zeeman-Doppler Imaging method \citep{donati09}. These are shown in Figure~\ref{fig:bfield maps}. The input for the Alfv\'en wave flux in our model is given in units of flux per unit magnetic field strength $S_\textrm{A}/B$. We refer to this as the `Alfv\'en wave flux-to-magnetic field ratio', for which we adopt a variety of values (see Section~\ref{sec:model results}). We fix the density and temperature at the base of the chromosphere as $2\times10^{10}$~cm$^{-3}$ and $5\times10^4$~K respectively in our models, the same as chosen by \citet{alvaradogomez20}. These values are also typically used for the Sun \citep[see][]{sokolov13}.

The corona is heated through a turbulent energy cascade in our model, which is produced by the reflection and dissipation of Alfv\'en waves in the local plasma environment. The wave dissipation rate depends on the transverse correlation length $L_\perp$, which in turn is proportional to $B^{-1/2}$ \citep{hollweg86}. In our model the transverse correlation length is given in terms of the proportionality constant $L_\perp \sqrt{B}$. We adopt the default value of $L_\perp \sqrt{B} = 1.5\times10^9$~cm~$\sqrt{\textrm{G}}$ for this \citep{vanderholst14}. We also include both collisional and collisionless heat conduction, radiative cooling, and stochastic heating in our models, as described in \citet{vanderholst14}. We fix the adiabatic index to be $5/3$.

Our computational grid is spherical and extends from the base of the chromosphere to 100 times the stellar radius $R_\star$. The maximum resolution in our grid is $\sim5\times10^{-4}$~$R_\star$. We also adaptively-resolve current sheet regions in our grid. The number of cells in our grid is $\sim6$~million. We iteratively solve the MHD equations until the model reaches a steady state. We take this to be when the mass-loss rate, open magnetic flux, and angular momentum-loss rate of the wind vary by less than 10 percent between iterations.

\section{Stellar wind environments of the M-dwarfs}
\label{sec:model results}


\subsection{Prox Cen}
\label{sec:model results - prox cen}

Using an Alfv\'en wave flux-to-magnetic field ratio of $S_\textrm{A}/B = 5\times10^{4}$~erg~s$^{-1}$~cm$^{-2}$~G$^{-1}$, we obtain a stellar wind with a mass-loss rate of $\dot{M}=0.25~\dot{M}_{\sun}$ for Prox Cen. This is in agreement with the upper limit obtained by \citet{wood01}. The rest of the free parameters in the model are as described in Section~\ref{sec:modelling}. Our calculated mass-loss rate is in line with other published works on the wind of Prox Cen, which used proxies for the surface magnetic field \citep{garraffo16, alvaradogomez20}.

\begin{figure}
\includegraphics[width = \columnwidth]{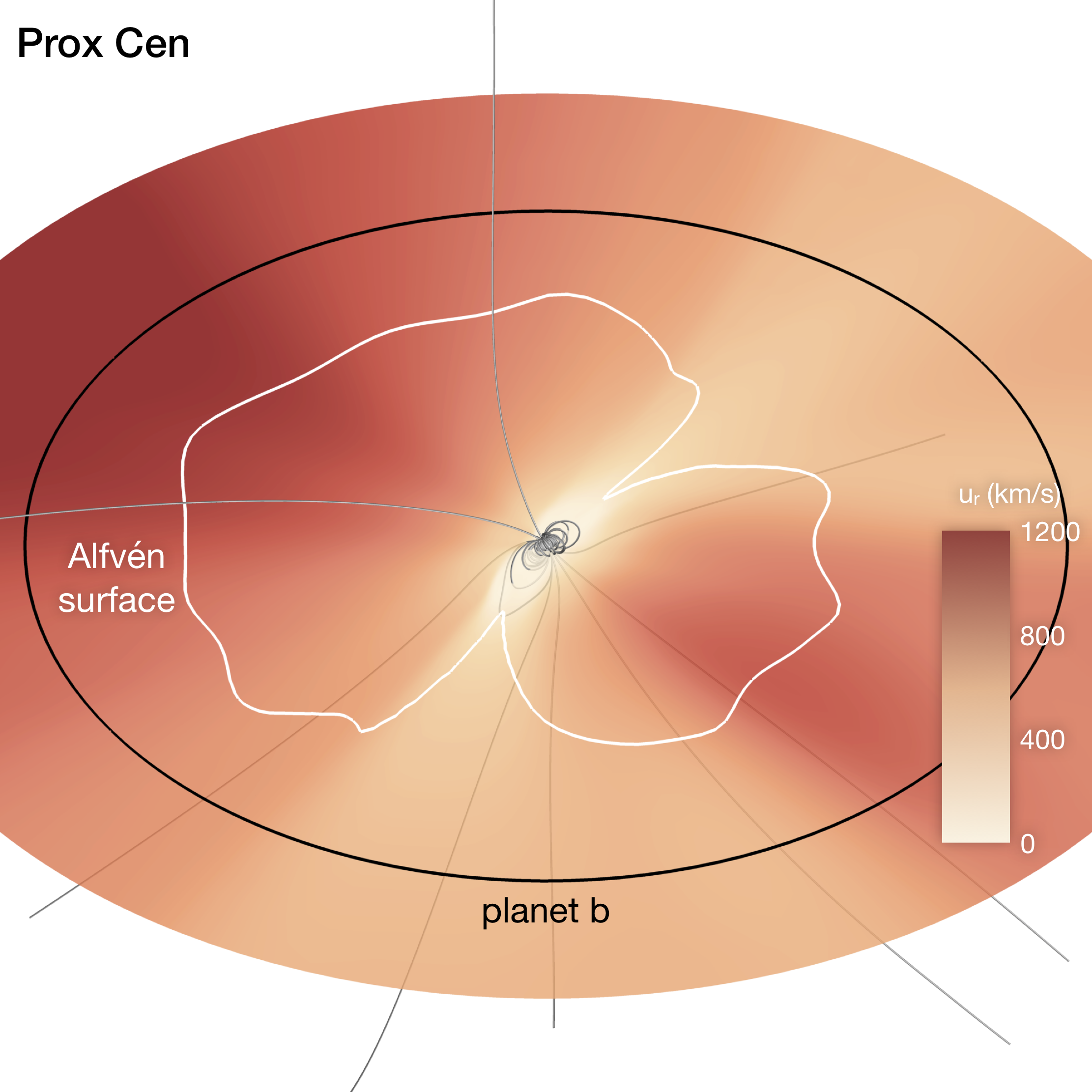}
\caption{Simulated stellar wind of Prox Cen. The grey lines illustrate the large-scale magnetic field of the star which is embedded in the wind. The orbit of Prox Cen b is shown as a black circle, and the white line corresponds to where the Alfv\'en Mach number $M_\textrm{A} = 1$ (see Equation~\ref{eq:mach number}). The contour in the orbital plane is coloured by the wind radial velocity ($u_r$).}
\label{fig:wind prox cen}
\end{figure}

Figure~\ref{fig:wind prox cen} shows a 3D view of the simulated wind for Prox Cen. We see that the planet orbits in a super-Alfv\'enic regime for the entirety of its orbit. Therefore, we do not expect any unipolar planet-induced radio emission from Prox Cen during the epoch of the observations from which the magnetic field map was reconstructed \citep[April - July 2017,][]{klein21}. Note that we assume that the orbital plane of the planet is perpendicular to the rotation axis of the star.

\subsection{AU Mic}
\label{sec:model results - au mic}

\begin{figure*}
\includegraphics[width = 0.495 \textwidth]{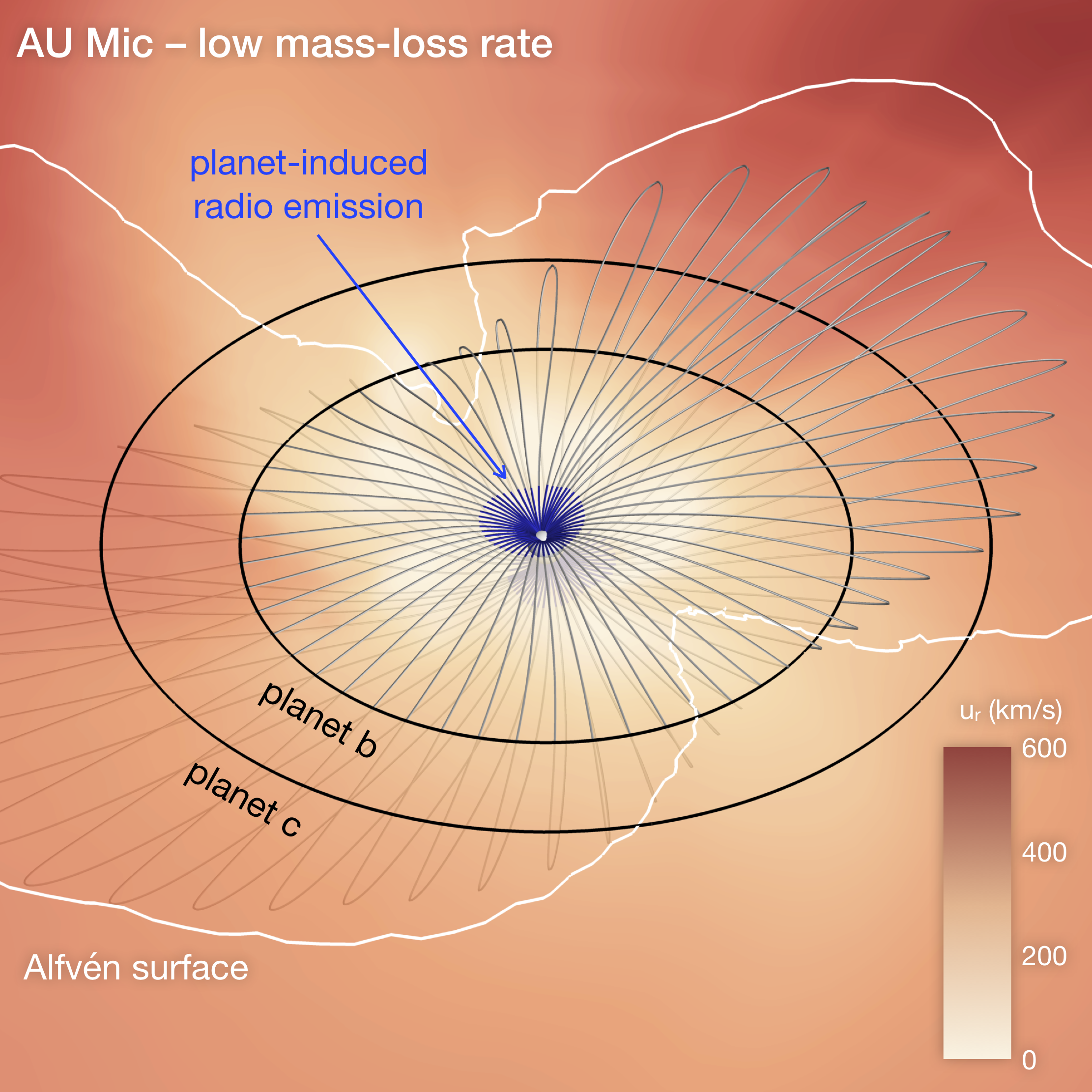}
\includegraphics[width = 0.495 \textwidth]{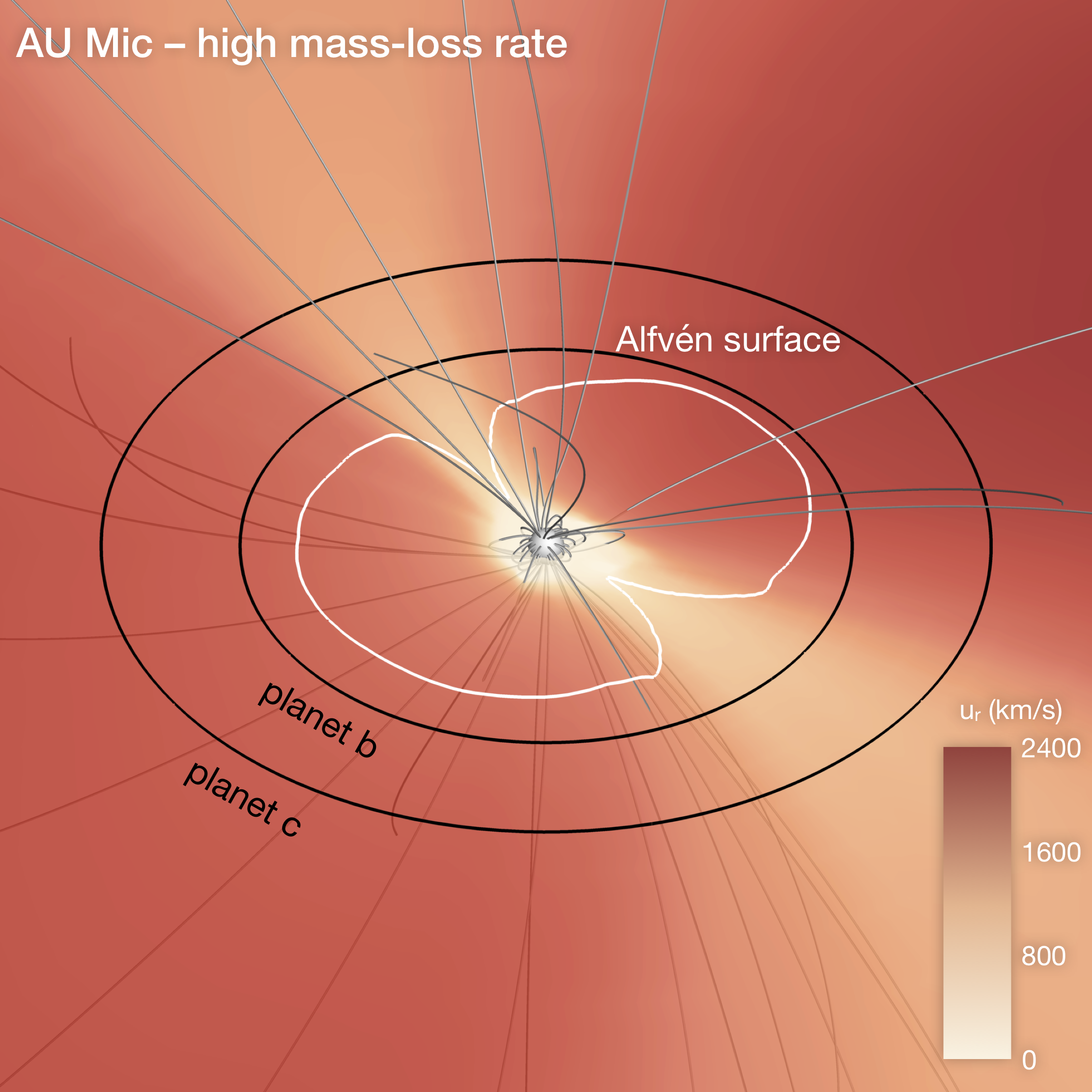}
\caption{\textit{Left}: Low $\dot{M}$ model of the stellar wind of AU Mic. The orbits of planets b and c are shown as black circles, and the white line corresponds to where the Alfv\'en Mach number $M_\textrm{A} = 1$ (see Equation~\ref{eq:mach number}). The contour in the orbital plane is coloured by the wind radial velocity ($u_r$). The grey lines show the stellar magnetic field lines that connect to the orbit of planet b. Each of these lines is a closed loop, and connects back to the star in both the Northern and Southern hemisphere. The blue shaded region of each line illustrates where planet b can induce the generation of radio emission via ECMI (see Equation~\ref{eq:plasma frequency}). Note that planet c can also induce generation, but for clarity we omit these field lines. \textit{Right}: High $\dot{M}$ model for AU Mic. Both planets b and c orbit in the super-Alfv\'enic regime in this scenario. Note that the magnetic field lines shown here do not connect to the orbit of either planet.}
\label{fig:wind au mic}
\end{figure*}

To investigate the two estimated mass-loss rates for AU Mic, we use two different values for the Alfv\'en wave flux-to-magnetic field ratio. Again, the values used for the rest of the free parameters in the model are given in Section~\ref{sec:modelling}. For a value of $S_\textrm{A}/B = 1.1\times10^{5}$~erg~s$^{-1}$~cm$^{-2}$~G$^{-1}$ \citep{sokolov13}, we obtain a stellar wind mass-loss rate of $27~\dot{M}_{\sun}$ for AU Mic. We refer to this as the `Low $\dot{M}$' model for the star. Increasing the Alfv\'en wave flux-to-magnetic field ratio to $S_\textrm{A}/B = 6\times10^{6}$~erg~s$^{-1}$~cm$^{-2}$~G$^{-1}$, we find that the mass-loss rate increases to a value of $590~\dot{M}_{\sun}$. We refer to this as the `High $\dot{M}$' model. Figure~\ref{fig:wind au mic} shows a 3D view of both stellar wind models for AU Mic. We take both planets to orbit perpendicular to the stellar rotation axis, as both planets transit the star with orbital inclinations of $\sim90^\circ$ \citep{martioli21}, with the orbit of planet b being aligned with the rotation axis of the star \citep{martioli20}.

In the case of the Low $\dot{M}$ model (left panel of Figure~\ref{fig:wind au mic}), we find that both planets b and c orbit in the sub-Alfv\'enic regime for the majority of their orbit. This means that the two planets could induce the generation of radio emission along the lines connecting the planet to the star \citep{turnpenney18, vedantham20}. For the High $\dot{M}$ model however, both planets are in the super-Alfv\'enic regime for their entire orbit (right panel of Figure~\ref{fig:wind au mic}), as the wind is much denser than the Low $\dot{M}$ model. As a result, we do not expect planet-induced radio emission from AU Mic in the case that it has a high mass-loss rate. 

We now estimate the maximum mass-loss rate for AU Mic at which the planets can induce radio emission. Using a least-squares method, we fit our data with a power law between the mass-loss rate and maximum size of the Alfv\'en surface in the orbital plane $R_\textrm{A, orb}^\textrm{max}$. For our data, we obtain a fit of $\dot{M} = 9.6\times10^5 (R_\textrm{A, orb}^\textrm{max})^{-2.6}~\dot{M}_\odot$, where $R_\textrm{A, orb}^\textrm{max}$ is in stellar radii. For both planets to be in sub-Alfvénic orbits at least partially, the maximum size of the Alfv\'en surface must be greater than the orbital distance of planet c. Substituting this orbital distance of $26.9~R_\star$ into our power law fit to our data, we find that AU Mic would need to have a mass-loss rate of $\lesssim190~\dot{M}_\odot$ in order for planet-induced radio emission to be generated in its corona. Detection of such emission would therefore allow us to place an upper limit of $\lesssim190~\dot{M}_{\sun}$ on the mass-loss rate of AU Mic. Note that in either case, radio emission could still be generated from the planet's own magnetosphere \citep[see][]{vidotto17, kavanagh19, kavanagh20}. We explore the scenario for the generation of planet-induced radio emission in AU Mic's corona further in Section~\ref{sec:radio emission au mic}.

\section{Planet-induced radio emission}


\subsection{AU Mic}
\label{sec:radio emission au mic}

Here, we describe our model we use to estimate the radio emission induced in the corona of AU Mic by planets b and c, in the scenario where the star has a mass-loss rate of $\lesssim190~\dot{M}_{\sun}$. We use our Low~$\dot{M}$ model data in our calculations. A planet is said to be in the sub-Alfv\'enic regime when the Alfv\'en Mach number is less than 1:
\begin{equation}
M_\textrm{A} = \frac{\Delta u}{u_\textrm{A}} < 1.
\label{eq:mach number}
\end{equation}
Here, $\Delta u$ is the relative velocity between the stellar wind velocity $u$ and planet orbital velocity $u_\textrm{p}$:
\begin{equation}
\boldsymbol{\Delta u} = \boldsymbol{u} - \boldsymbol{u_\textrm{p}}.
\end{equation}
The orbital velocity of the planet is $\boldsymbol{u_\textrm{p}} = \sqrt{GM_\star/a}~\boldsymbol{\hat{\phi}}$, where $G$ is the gravitational constant, $M_{\star}$ is the stellar mass, $a$ is the orbital distance, and $\boldsymbol{\hat{\phi}}$ is the azimuthal angle. In equation~\ref{eq:mach number}, $u_\textrm{A}$ is the Alfv\'en velocity of the stellar wind at the planet:
\begin{equation}
u_\textrm{A} = \frac{B}{\sqrt{4\pi\rho}},
\end{equation}
where $B$ and $\rho$ are the stellar wind magnetic field strength and density respectively.

The power of the Alfv\'en waves produced by a planet orbiting in the sub-Alfv\'enic regime is \citep{saur13}:
\begin{equation}
P = \pi^{1/2} R^2 B \rho^{1/2} \Delta u^2 \sin^2 \theta,
\label{eq:radio power}
\end{equation}
where $R$ is the radius of the obstacle, and $\theta$ is the angle between the vectors $\boldsymbol{B}$ and $\boldsymbol{\Delta u}$. Assuming the planet is unmagnetised, we use the planetary radius $R_\textrm{p}$ for the obstacle radius. The waves travel towards the star along the magnetic field line connecting it to the planet, and a fraction of their energy is converted into ECMI emission. The emission occurs at the local cyclotron frequency \citep{turnpenney18}: 
\begin{equation}
\nu = 2.8\times10^{-3} \frac{B}{\textrm{1 G}}~\textrm{GHz}.
\label{eq:cyclotron frequency}
\end{equation}
We assume emission can be generated everywhere along the magnetic field line connecting the planet to the star, provided the cyclotron frequency exceeds the plasma frequency:
\begin{equation}
\nu > \nu_\textrm{p} = 9\times10^{-6}\sqrt{\frac{n_\textrm{e}}{\textrm{1 cm}^{-3}}}~\textrm{GHz}.
\label{eq:plasma frequency}
\end{equation}
Here $n_\textrm{e}$ is the electron number density. 

We trace each magnetic field line that connects to the planet's orbit back to the point where it connects to the surface of AU Mic. Both planets orbit through closed-field regions of the star's large-scale magnetic field, which predominantly resembles an aligned dipole \citep{klein21}. This means that ECMI emission can be generated in both the Northern and Southern hemisphere of AU Mic. Checking the condition in Equation~\ref{eq:plasma frequency} is satisfied, we extract the magnetic field strength along the line and compute the cyclotron frequency with Equation~\ref{eq:cyclotron frequency}. These regions are highlighted in blue in the left panel of Figure~\ref{fig:wind au mic}. We then compute the flux density observed at a distance $d$ from the system as
\begin{equation}
F_\nu = \frac{\varepsilon P}{\Omega d^2 \nu}.
\label{eq:flux}
\end{equation}
Here $\varepsilon$ is the fraction of the wave energy from the interaction which is converted into ECMI emission. Observations of the Jupiter-Io interaction indicate that $\varepsilon \approx 0.01$ \citep{turnpenney18}, which we adopt here. $\Omega$ is the solid angle of the emission beam. Again, we adopt the observed value for the Jupiter-Io interaction of $\Omega = 1.6$~sr \citep{zarka04}.

\begin{figure*}
\centering
\includegraphics[width = \textwidth]{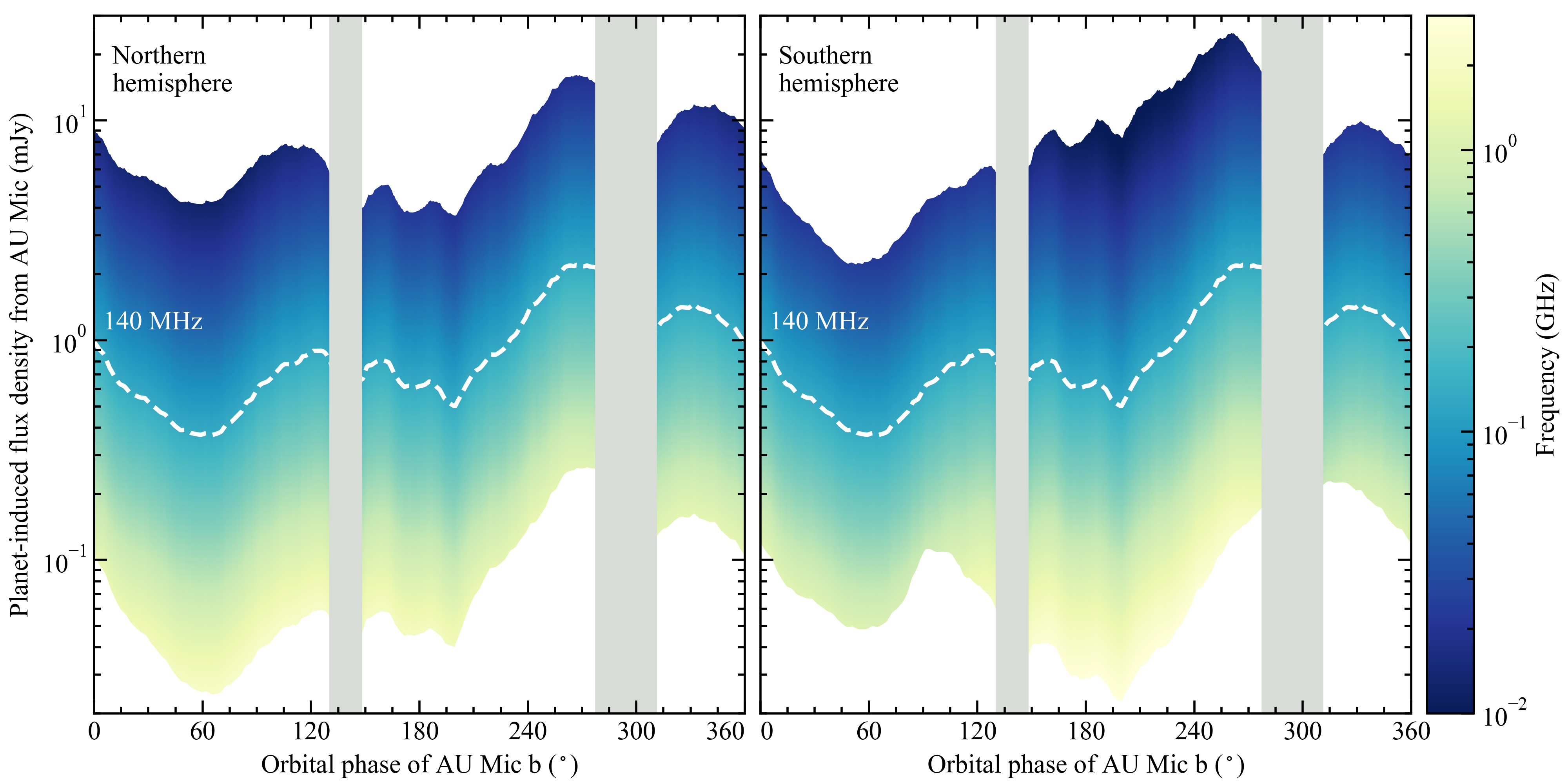}
\caption{Radio spectrum of AU Mic induced by planet b in the Northern and Southern hemispheres of the star's corona (left and right panels respectively). Emission generated at 140~MHz is highlighted with a white dashed line. This is the middle frequency of the observing band at which radio emission was recently detected from the M~dwarf GJ~1151 by \citet{vedantham20}, which is suspected of being induced by an orbiting planet. The grey shaded areas illustrate the region where the orbit of planet is in the super-Alfv\'enic regime. No emission can generate in these regions (see Equation~\ref{eq:plasma frequency}).}
\label{fig:spec au mic}
\end{figure*}

We show the flux density induced by planet b in the Northern and Southern hemispheres of AU Mic as a function of its orbital phase in Figure~\ref{fig:spec au mic}. The flux density is colour-coded with the frequency of the emission. We find that planet b can induce emission from $\sim10$~MHz -- 3~GHz in both hemispheres, with flux densities ranging from $\sim10^{-2}$ -- 10~mJy. Higher flux densities correspond to lower frequencies (see Equation~\ref{eq:flux}). Note that planet c also can induce radio emission in this frequency range, albeit with flux densities that are around an order of magnitude lower. This is due to the term $B\rho^{1/2}$ in Equation~\ref{eq:radio power} dominating at larger orbital distances, as the density and magnetic field strength drop off radially. As the planets orbit with different rotation periods however, they may produce distinct radio signals at different frequencies and periodicities that could be detected.

In Figure~\ref{fig:spec au mic} we also highlight the emission that is generated at 140~MHz, which corresponds to the middle of the frequency range of 120 -- 160~MHz at which some M~dwarfs have recently been detected \citep[][Callingham et al., submitted]{vedantham20}. \citet{vedantham20} suggested that their observations of emission from the M dwarf GJ~1151 may be generated by a planet orbiting in the sub-Alfv\'enic regime. At 140~MHz, our results bear a strong resemblance to the observations of GJ~1151: both have flux densities of about 1~mJy which exhibit temporal variations. It would be very useful to obtain radio and near-simultaneous spectropolarimetric observations of M~dwarfs similar to these systems so that this scenario could be explored further. 

Time-varying nonthermal radio emission from AU Mic at 1.5~GHz was reported by \citet{cox85}, with flux densities ranging from 2 -- 3~mJy. Compared to our predicted fluxes induced by planet b at this frequency range, the observed flux densities are about an order of magnitude larger. If this emission was induced by either planet, it would require a much higher radio power than we predict. This could occur if the wind properties were enhanced in response to an increase in the magnetic field strength of AU Mic. Alternatively, if the radius of the obstacle were $R \approx 3.2~R_\textrm{p}$, the radio power estimated using Equation~\ref{eq:radio power} would increase by an order of magnitude. We estimate that this would be the size of the magnetopause of planet b if it had a dipolar field strength of $\sim6$~G \citep[see][]{vidotto17, kavanagh19}. Therefore, not only could detection of planet-induced radio emission allow us to constrain the mass-loss rate of the host star, but also the magnetic field strength of the planet itself, if it is magnetised. As there is no polarisation information reported by \citet{cox85} however, we unfortunately cannot assess the origin of this emission this further. Future observations with the VLA, MWA, or upcoming SKA radio telescope would certainly be beneficial to search for radio signals in the frequency range at which we predict AU Mic could emit planet-induced radio emission. A derivable quantity such as the brightness temperature of the emission could allow for detected polarised emission indicative of being planet-induced to be distinguished from flaring emission, as has been illustrated by \citet{vedantham20}.

\subsection{Prox Cen}
\label{sec:radio emission prox cen}

Radio emission from 1.1 -- 3.1~GHz was recently detected from Prox Cen by \citet{pereztorres21}. These observations occurred during April 2017, at same epoch the spectropolarimetric observations began for Prox Cen \citep{klein21}. \citet{pereztorres21} suggested that this emission could be indicative of star-planet interactions. If this is the case, it would require the planet to be orbiting in the sub-Alfv\'enic regime. Our wind model of Prox Cen however, which has a mass-loss rate in agreement with the upper limit obtained by \citet{wood01}, shows that planet b orbits in the super-Alfv\'enic regime for the entirety of its orbit. In order for the planet to induce radio emission from the corona of Prox Cen, the Alfv\'en surface would need to extend beyond the planet's orbit. This would require the mass-loss rate of the star to be significantly lower than the value we obtained of $0.25~\dot{M}_{\sun}$.

To investigate this further, we ran two additional stellar wind models for Prox Cen, with Alfv\'en wave flux-to-magnetic field ratios of $S_\textrm{A}/B = 1.1\times10^4$~erg~s$^{-1}$~cm$^{-2}$~G$^{-1}$ and $1.1\times10^5$~erg~s$^{-1}$~cm$^{-2}$~G$^{-1}$. We obtain mass-loss rates of $0.05~\dot{M}_{\sun}$ and $0.57~\dot{M}_{\sun}$ respectively for these two simulations. In the case of the lower mass-loss rate model, we find that the planet still orbits entirely in a super-Alfv\'enic regime. Naturally, this is also the case for the model $0.57~\dot{M}_{\sun}$, which has a denser wind than our model presented in Section~\ref{sec:model results - prox cen}.

We now use these results, combined with those from Section~\ref{sec:model results - prox cen}, to estimate the Alfv\'en wave flux-to-magnetic field ratio and mass-loss rate needed to place Prox Cen b in a sub-Alfv\'enic orbit. We first perform a least-squares power law fit between the Alfv\'en wave flux-to-magnetic field ratio $S_\textrm{A}/B$ and mass-loss rate $\dot{M}$, and find that $\dot{M} = 2.6\times10^{-6} (S_\textrm{A}/B)^{1.06}~\dot{M}_{\sun}$ for Prox Cen, where $S_\textrm{A}/B$ is in erg~s$^{-1}$~cm$^{-2}$~G$^{-1}$. The effects of changing $S_\textrm{A}/B$ on the solar wind mass-loss rate were investigated by \citet{borosaikia20} using the AWSoM model. For comparison, we fit the same power law to their results, and find that during solar minimum $\dot{M} \propto (S_\textrm{A}/B)^{1.37}$, and during solar maximum $\dot{M} \propto (S_\textrm{A}/B)^{1.42}$ (using the $l_\textrm{max}=5$ data in their Tables F.1 and F.2).

\begin{figure}
\centering
\includegraphics[width = \columnwidth]{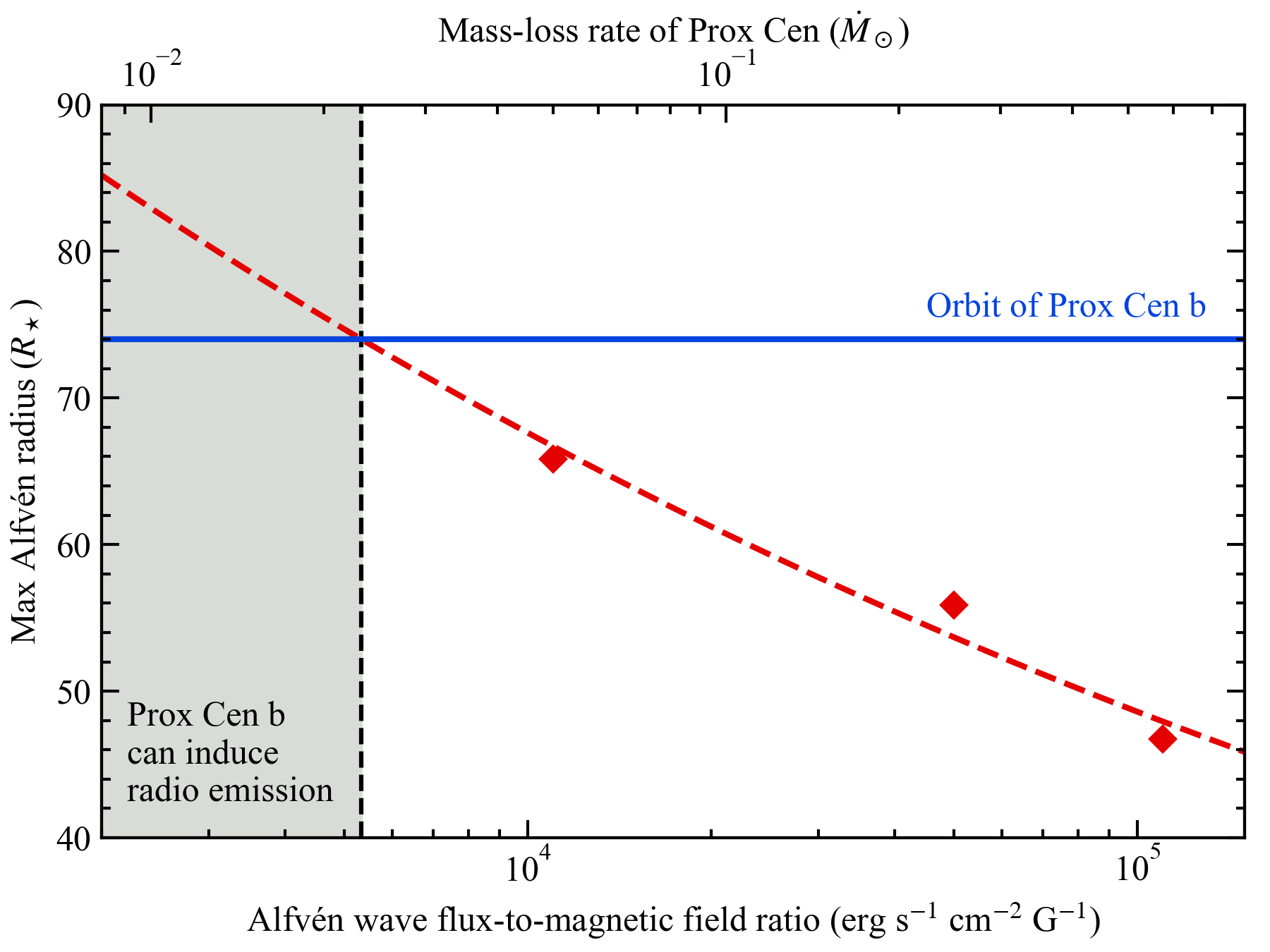}
\caption{Average radius of the Alfv\'en surface in the orbital plane of Prox Cen b vs.~the Alfv\'en wave flux-to-magnetic field ratio (red diamonds). The upper $x$-axis is scaled accordingly based on the mass-loss rate of each wind simulation. The red dashed line illustrates our least-squares power law fit to the datapoints. We extrapolate this fit to estimate the Alfv\'en wave flux-to-magnetic field ratio required for the Alfv\'en surface to exceed the orbital distance of the planet (blue horizontal line). If this condition is satisfied, the planet can induce the generation of radio emission in the corona of the Prox Cen. This region is shown in grey on the left-hand side. We find that an Alfv\'en wave flux-to-magnetic field ratio of $\leq 2.7\times10^{-3}$~erg~s$^{-1}$~cm$^{-2}$~G$^{-1}$ is needed for the planet to orbit inside the Alfv\'en surface. This corresponds to mass-loss rate of $\leq 1.1\times10^{-2}~\dot{M}_{\sun}$ for Prox Cen.}
\label{fig:wind prox cen fit}
\end{figure}

For Prox Cen, we compute the maximum radius of the Alfv\'en surface $R_\textrm{A, orb}^\textrm{max}$ in the orbital plane of the planet for each plot, and plot it with its corresponding Alfv\'en wave flux-to-magnetic field ratio and mass-loss rate in Figure~\ref{fig:wind prox cen fit}. Again, performing a least-squares power law fit between $R_\textrm{A, orb}^\textrm{max}$ and $S_\textrm{A}/B$, we estimate that an Alfv\'en wave flux-to-magnetic field ratio of $< 5.3\times10^3$~erg~s$^{-1}$~cm$^{-2}$~G$^{-1}$ is required in order to place the plane in a sub-Alfv\'enic orbit. This corresponds to a stellar wind mass-loss rate of $< 2.3\times10^{-2}~\dot{M}_{\sun}$, which is an order of magnitude lower than the upper limit obtained by \citet{wood01}.

\section{Discussion}

There are some caveats to our planet-induced radio emission model. Firstly, we assume that the emission is always visible to the observer. However, it is well-established from observations of the Jupiter-Io interaction that the emission generated in this interaction is beamed in a hollow cone \citep[see][]{zarka04}. As a result, if the emission can be generated it would likely be beamed in and out of the line of sight towards the system. Therefore, our results represent the best-case scenario where we always see the emission. We also do not account for the relative rotational motion between the star and planet, which would effect which stellar magnetic field line connects to the planet at any given time. Doing so would allow for us to estimate the temporal variations one would see observing the system, if this emission were detectable.

We also note that the higher radio fluxes we estimate for AU Mic in the case of a low mass-loss rate correspond to lower frequencies. However, as was shown by \citet{kavanagh20}, low frequency radio emission could suffer from significant attenuation from the stellar wind. We computed the thermal spectrum of AU Mic's low mass-loss rate wind using the model developed by \citet{ofionnagain19, ofionnagain19-erratum}, and found that the wind is optically thick below $\sim0.1$~GHz. This means that a significant amount of the low frequency planet-induced emission ($<0.1$~GHz) we predict could be absorbed by the stellar wind. Note that we compute a peak flux density of $\sim0.5$~$\mu$Jy for the thermal wind spectrum of AU Mic, which is 2 -- 3 orders of magnitude smaller than the predicted planet-induced emission.

While we do not expect planet-induced radio emission from Prox Cen or AU Mic in the case that it has a high mass-loss rate, radio emission could still be generated in the magnetospheres of the orbiting planets \citep[see][]{vidotto19}. For a low mass-loss rate however, AU Mic seems to be promising candidate for detectable planet-induced radio emission. Simultaneous spectropolarimetric and radio data of the system would be very complimentary to investigate this further, which would allow us to assess the stellar wind environment of AU Mic at the time of the radio observations. UV observations of the planetary transits could also be used in parallel, which have been shown to vary depending on the mass-loss rate of the host star \citep{carolan20}. In a broader context, the methods presented in the paper could be applied to a wide range of other suitable exoplanetary systems which may be capable of generating planet-induced radio emission.


\section{Conclusions}

In this paper, we presented Alfv\'en wave-driven stellar wind models for the two active M dwarfs Prox Cen and AU Mic, both of which are hosts to planets. We used our stellar wind models to investigate whether the orbiting planets could orbit sub-Alfv\'enically and induce radio emission in the coronae of their host stars, through a Jupiter-Io-like interaction. For Prox Cen, we constrained our stellar wind model using the upper limit for the mass-loss rate of $0.2~\dot{M}_{\sun}$ obtained by \citet{wood01}. We found that for this mass-loss rate, the planet orbits in the super-Alfv\'enic regime for its entire orbit. As a result, we do not expect that the planet could induce radio emission from the corona of the host star. By performing additional stellar wind simulations and extrapolating the results, we estimate that Prox Cen would need to have a mass-loss rate of $\leq 1.1\times10^{-2}~\dot{M}_{\sun}$ in order for the planet to orbit sub-Alfv\'enically. In other words, Prox Cen would need to have a mass-loss rate that is 20 times lower than the expected upper limit in order for the orbiting planet to induce radio emission in its corona.

For AU Mic, we investigated two scenarios, where the star has a low and high mass-loss rate. By adjusting the flux of Alfv\'en waves propagating out of the photosphere in our models, we obtained mass-loss rates of $27~\dot{M}_{\sun}$ and $590~\dot{M}_{\sun}$ respectively for the star. In the case of the low mass-loss rate ($27~\dot{M}_{\sun}$), we found that both planets b and c orbit sub-Alfv\'enically for the majority of their orbits. This means that both planets could induce the generation of radio emission in the corona of AU Mic. We showed that planet b could induce time-varying radio emission from $\sim10$~MHz -- 3~GHz, with a peak flux density of $\sim10$~mJy. The radio emission we predict also bears a striking resemblance to that reported for the M~dwarf GJ~1151 by \citet{vedantham20}, which is suspected of be induced by a terrestrial planet in a 1 -- 5 day orbit. Planet c could also induce radio emission in this frequency range, but with flux densities that are about an order of magnitude lower, and for a smaller fraction of its orbit. Due to the differences in orbital periods, both planets b and c could therefore produce radio emission distinct from one another. Comparing our estimated planet-induced radio emission for AU Mic to observations reported by \citet{cox85}, we found that planet b could be magnetised, with a field strength of $\sim6$~G. However, as there is no polarisation information reported by \citet{cox85}, we cannot assess further if this emission is indicative of being induced by the planets. 

Our results illustrate that detection of planet-induced radio emission from AU Mic could allow us to constrain both the mass-loss rate of the star, as well as the magnetic field strength of the planet that induces the emission.


\section*{Acknowledgements}

We thank the anonymous referee for their comments and suggestions. We also thank Dr.~Joe Callingham, Dr.~Sebastian Pineda, Prof.~Harish Vedantham, and Prof.~Jackie Villadsen for their insightful discussions. RDK acknowledges funding received from the Irish Research Council (IRC) through the Government of Ireland Postgraduate Scholarship Programme. AAV acknowledges funding from the European Research Council (ERC) under the European Union's Horizon 2020 research and innovation programme (grant agreement No 817540, ASTROFLOW). BK acknowledges funding from the European Research Council under the European Union's Horizon 2020 research and innovation programme (grant agreement No 865624, GPRV). MMJ acknowledges support from STFC consolidated grant number ST/R000824/1. JFD and BK acknowledge the ERC for grant agreement No 740651, NewWorlds. D\'{O}F acknowledges funding from the IRC Government of Ireland Postdoctoral Fellowship Programme. We acknowledge the Irish Centre for High-End Computing (ICHEC) for providing the computational facilities used to perform the simulations published in this work.


\section*{Data availability}

The data presented in this paper will be shared on reasonable request to the corresponding author.


\bibliographystyle{mnras}
\bibliography{bibliography}

\begin{thebibliography}{}
\makeatletter
\relax
\def\mn@urlcharsother{\let\do\@makeother \do\$\do\&\do\#\do\^\do\_\do\%\do\~}
\def\mn@doi{\begingroup\mn@urlcharsother \@ifnextchar [ {\mn@doi@}
  {\mn@doi@[]}}
\def\mn@doi@[#1]#2{\def\@tempa{#1}\ifx\@tempa\@empty \href
  {http://dx.doi.org/#2} {doi:#2}\else \href {http://dx.doi.org/#2} {#1}\fi
  \endgroup}
\def\mn@eprint#1#2{\mn@eprint@#1:#2::\@nil}
\def\mn@eprint@arXiv#1{\href {http://arxiv.org/abs/#1} {{\tt arXiv:#1}}}
\def\mn@eprint@dblp#1{\href {http://dblp.uni-trier.de/rec/bibtex/#1.xml}
  {dblp:#1}}
\def\mn@eprint@#1:#2:#3:#4\@nil{\def\@tempa {#1}\def\@tempb {#2}\def\@tempc
  {#3}\ifx \@tempc \@empty \let \@tempc \@tempb \let \@tempb \@tempa \fi \ifx
  \@tempb \@empty \def\@tempb {arXiv}\fi \@ifundefined
  {mn@eprint@\@tempb}{\@tempb:\@tempc}{\expandafter \expandafter \csname
  mn@eprint@\@tempb\endcsname \expandafter{\@tempc}}}

\bibitem[\protect\citeauthoryear{{Alvarado-G{\'o}mez}
  et~al.,}{{Alvarado-G{\'o}mez} et~al.}{2020}]{alvaradogomez20}
{Alvarado-G{\'o}mez} J.~D.,  et~al., 2020, \mn@doi [\apj]
  {10.3847/1538-4357/ab88a3}, \href
  {https://ui.adsabs.harvard.edu/abs/2020ApJ...895...47A} {895, 47}

\bibitem[\protect\citeauthoryear{{Anglada-Escud{\'e}}
  et~al.,}{{Anglada-Escud{\'e}} et~al.}{2016}]{angladaescude16}
{Anglada-Escud{\'e}} G.,  et~al., 2016, \mn@doi [\nat] {10.1038/nature19106},
  \href {https://ui.adsabs.harvard.edu/abs/2016Natur.536..437A} {536, 437}

\bibitem[\protect\citeauthoryear{{Bastian}, {Dulk}  \& {Leblanc}}{{Bastian}
  et~al.}{2000}]{bastian00}
{Bastian} T.~S.,  {Dulk} G.~A.,   {Leblanc} Y.,  2000, \mn@doi [\apj]
  {10.1086/317864}, \href
  {https://ui.adsabs.harvard.edu/abs/2000ApJ...545.1058B} {545, 1058}

\bibitem[\protect\citeauthoryear{{Boro Saikia}, {Jin}, {Johnstone},
  {L{\"u}ftinger}, {G{\"u}del}, {Airapetian}, {Kislyakova}  \& {Folsom}}{{Boro
  Saikia} et~al.}{2020}]{borosaikia20}
{Boro Saikia} S.,  {Jin} M.,  {Johnstone} C.~P.,  {L{\"u}ftinger} T.,
  {G{\"u}del} M.,  {Airapetian} V.~S.,  {Kislyakova} K.~G.,   {Folsom} C.~P.,
  2020, \mn@doi [\aap] {10.1051/0004-6361/201937107}, \href
  {https://ui.adsabs.harvard.edu/abs/2020A&A...635A.178B} {635, A178}

\bibitem[\protect\citeauthoryear{{Carolan}, {Vidotto}, {Plavchan}, {Villarreal
  D'Angelo}  \& {Hazra}}{{Carolan} et~al.}{2020}]{carolan20}
{Carolan} S.,  {Vidotto} A.~A.,  {Plavchan} P.,  {Villarreal D'Angelo} C.,
  {Hazra} G.,  2020, \mn@doi [\mnras] {10.1093/mnrasl/slaa127}, \href
  {https://ui.adsabs.harvard.edu/abs/2020MNRAS.498L..53C} {498, L53}

\bibitem[\protect\citeauthoryear{{Chiang} \& {Fung}}{{Chiang} \&
  {Fung}}{2017}]{chiang17}
{Chiang} E.,  {Fung} J.,  2017, \mn@doi [\apj] {10.3847/1538-4357/aa89e6},
  \href {https://ui.adsabs.harvard.edu/abs/2017ApJ...848....4C} {848, 4}

\bibitem[\protect\citeauthoryear{Cox \& Gibson}{Cox \& Gibson}{1985}]{cox85}
Cox J.~J.,  Gibson D.~M.,  1985, in Hjellming R.~M.,  Gibson D.~M.,  eds, Radio
  Stars. Springer Netherlands, Dordrecht, pp 233--236

\bibitem[\protect\citeauthoryear{{Donati} \& {Landstreet}}{{Donati} \&
  {Landstreet}}{2009}]{donati09}
{Donati} J.~F.,  {Landstreet} J.~D.,  2009, \mn@doi [\araa]
  {10.1146/annurev-astro-082708-101833}, \href
  {https://ui.adsabs.harvard.edu/abs/2009ARA&A..47..333D} {47, 333}

\bibitem[\protect\citeauthoryear{{Donati} et~al.,}{{Donati}
  et~al.}{2008}]{donati08}
{Donati} J.~F.,  et~al., 2008, \mn@doi [\mnras]
  {10.1111/j.1365-2966.2008.13799.x}, \href
  {https://ui.adsabs.harvard.edu/abs/2008MNRAS.390..545D} {390, 545}

\bibitem[\protect\citeauthoryear{{Garraffo}, {Drake}  \& {Cohen}}{{Garraffo}
  et~al.}{2016}]{garraffo16}
{Garraffo} C.,  {Drake} J.~J.,   {Cohen} O.,  2016, \mn@doi [\apjl]
  {10.3847/2041-8205/833/1/L4}, \href
  {https://ui.adsabs.harvard.edu/abs/2016ApJ...833L...4G} {833, L4}

\bibitem[\protect\citeauthoryear{{Garraffo}, {Drake}, {Cohen},
  {Alvarado-G{\'o}mez}  \& {Moschou}}{{Garraffo} et~al.}{2017}]{garraffo17}
{Garraffo} C.,  {Drake} J.~J.,  {Cohen} O.,  {Alvarado-G{\'o}mez} J.~D.,
  {Moschou} S.~P.,  2017, \mn@doi [\apjl] {10.3847/2041-8213/aa79ed}, \href
  {https://ui.adsabs.harvard.edu/abs/2017ApJ...843L..33G} {843, L33}

\bibitem[\protect\citeauthoryear{{Grie{\ss}meier}, {Zarka}  \&
  {Spreeuw}}{{Grie{\ss}meier} et~al.}{2007}]{griessmeier07}
{Grie{\ss}meier} J.~M.,  {Zarka} P.,   {Spreeuw} H.,  2007, \mn@doi [\aap]
  {10.1051/0004-6361:20077397}, \href
  {https://ui.adsabs.harvard.edu/abs/2007A&A...475..359G} {475, 359}

\bibitem[\protect\citeauthoryear{{Hollweg}}{{Hollweg}}{1986}]{hollweg86}
{Hollweg} J.~V.,  1986, \mn@doi [\jgr] {10.1029/JA091iA04p04111}, \href
  {https://ui.adsabs.harvard.edu/abs/1986JGR....91.4111H} {91, 4111}

\bibitem[\protect\citeauthoryear{{Ip}, {Kopp}  \& {Hu}}{{Ip}
  et~al.}{2004}]{ip04}
{Ip} W.-H.,  {Kopp} A.,   {Hu} J.-H.,  2004, \mn@doi [\apjl] {10.1086/382274},
  \href {https://ui.adsabs.harvard.edu/abs/2004ApJ...602L..53I} {602, L53}

\bibitem[\protect\citeauthoryear{{Jin}, {Manchester}, {van der Holst},
  {Sokolov}, {T{\'o}th}, {Vourlidas}, {de Koning}  \& {Gombosi}}{{Jin}
  et~al.}{2017}]{jin17}
{Jin} M.,  {Manchester} W.~B.,  {van der Holst} B.,  {Sokolov} I.,  {T{\'o}th}
  G.,  {Vourlidas} A.,  {de Koning} C.~A.,   {Gombosi} T.~I.,  2017, \mn@doi
  [\apj] {10.3847/1538-4357/834/2/172}, \href
  {https://ui.adsabs.harvard.edu/abs/2017ApJ...834..172J} {834, 172}

\bibitem[\protect\citeauthoryear{{Kavanagh} \& {Vidotto}}{{Kavanagh} \&
  {Vidotto}}{2020}]{kavanagh20}
{Kavanagh} R.~D.,  {Vidotto} A.~A.,  2020, \mn@doi [\mnras]
  {10.1093/mnras/staa422}, \href
  {https://ui.adsabs.harvard.edu/abs/2020MNRAS.493.1492K} {493, 1492}

\bibitem[\protect\citeauthoryear{{Kavanagh} et~al.,}{{Kavanagh}
  et~al.}{2019}]{kavanagh19}
{Kavanagh} R.~D.,  et~al., 2019, \mn@doi [\mnras] {10.1093/mnras/stz655}, \href
  {https://ui.adsabs.harvard.edu/abs/2019MNRAS.485.4529K} {485, 4529}

\bibitem[\protect\citeauthoryear{{Klein} et~al.,}{{Klein}
  et~al.}{2020}]{klein20}
{Klein} B.,  et~al., 2020, \mn@doi [\mnras] {10.1093/mnras/staa3702}, \href
  {https://ui.adsabs.harvard.edu/abs/2020MNRAS.tmp.3555K} {}

\bibitem[\protect\citeauthoryear{{Klein}, {Donati}, {H{\'e}brard}, {Zaire},
  {Folsom}, {Morin}, {Delfosse}  \& {Bonfils}}{{Klein} et~al.}{2021}]{klein21}
{Klein} B.,  {Donati} J.-F.,  {H{\'e}brard} {\'E}.~M.,  {Zaire} B.,  {Folsom}
  C.~P.,  {Morin} J.,  {Delfosse} X.,   {Bonfils} X.,  2021, \mn@doi [\mnras]
  {10.1093/mnras/staa3396}, \href
  {https://ui.adsabs.harvard.edu/abs/2021MNRAS.500.1844K} {500, 1844}

\bibitem[\protect\citeauthoryear{{Lanza}}{{Lanza}}{2012}]{lanza12}
{Lanza} A.~F.,  2012, \mn@doi [\aap] {10.1051/0004-6361/201219002}, \href
  {https://ui.adsabs.harvard.edu/abs/2012A&A...544A..23L} {544, A23}

\bibitem[\protect\citeauthoryear{{Lazio}, {Carmichael}, {Clark}, {Elkins},
  {Gudmundsen}, {Mott}, {Szwajkowski}  \& {Hennig}}{{Lazio}
  et~al.}{2010}]{lazio10}
{Lazio} T. J.~W.,  {Carmichael} S.,  {Clark} J.,  {Elkins} E.,  {Gudmundsen}
  P.,  {Mott} Z.,  {Szwajkowski} M.,   {Hennig} L.~A.,  2010, \mn@doi [\aj]
  {10.1088/0004-6256/139/1/96}, \href
  {https://ui.adsabs.harvard.edu/abs/2010AJ....139...96L} {139, 96}

\bibitem[\protect\citeauthoryear{{Lecavelier des Etangs}, {Sirothia},
  {Gopal-Krishna}  \& {Zarka}}{{Lecavelier des Etangs}
  et~al.}{2013}]{lecavelier13}
{Lecavelier des Etangs} A.,  {Sirothia} S.~K.,  {Gopal-Krishna}  {Zarka} P.,
  2013, \mn@doi [\aap] {10.1051/0004-6361/201219789}, \href
  {https://ui.adsabs.harvard.edu/abs/2013A&A...552A..65L} {552, A65}

\bibitem[\protect\citeauthoryear{{Mahadevan} et~al.,}{{Mahadevan}
  et~al.}{2021}]{mahadevan21}
{Mahadevan} S.,  et~al., 2021, arXiv e-prints, \href
  {https://ui.adsabs.harvard.edu/abs/2021arXiv210202233M} {p. arXiv:2102.02233}

\bibitem[\protect\citeauthoryear{{Martioli}, {H{\'e}brard}, {Correia}, {Laskar}
   \& {Lecavelier des Etangs}}{{Martioli} et~al.}{2020a}]{martioli21}
{Martioli} E.,  {H{\'e}brard} G.,  {Correia} A.~C.~M.,  {Laskar} J.,
  {Lecavelier des Etangs} A.,  2020a, arXiv e-prints, \href
  {https://ui.adsabs.harvard.edu/abs/2020arXiv201213238M} {p. arXiv:2012.13238}

\bibitem[\protect\citeauthoryear{{Martioli} et~al.,}{{Martioli}
  et~al.}{2020b}]{martioli20}
{Martioli} E.,  et~al., 2020b, \mn@doi [\aap] {10.1051/0004-6361/202038695},
  \href {https://ui.adsabs.harvard.edu/abs/2020A&A...641L...1M} {641, L1}

\bibitem[\protect\citeauthoryear{{McIvor}, {Jardine}  \& {Holzwarth}}{{McIvor}
  et~al.}{2006}]{mcivor06}
{McIvor} T.,  {Jardine} M.,   {Holzwarth} V.,  2006, \mn@doi [\mnras]
  {10.1111/j.1745-3933.2005.00098.x}, \href
  {https://ui.adsabs.harvard.edu/abs/2006MNRAS.367L...1M} {367, L1}

\bibitem[\protect\citeauthoryear{{Morin}, {Donati}, {Petit}, {Delfosse},
  {Forveille}  \& {Jardine}}{{Morin} et~al.}{2010}]{morin10}
{Morin} J.,  {Donati} J.~F.,  {Petit} P.,  {Delfosse} X.,  {Forveille} T.,
  {Jardine} M.~M.,  2010, \mn@doi [\mnras] {10.1111/j.1365-2966.2010.17101.x},
  \href {https://ui.adsabs.harvard.edu/abs/2010MNRAS.407.2269M} {407, 2269}

\bibitem[\protect\citeauthoryear{{Narang} et~al.,}{{Narang}
  et~al.}{2021}]{narang21}
{Narang} M.,  et~al., 2021, \mn@doi [\mnras] {10.1093/mnras/staa3565}, \href
  {https://ui.adsabs.harvard.edu/abs/2021MNRAS.500.4818N} {500, 4818}

\bibitem[\protect\citeauthoryear{{Neubauer}}{{Neubauer}}{1980}]{neubauer80}
{Neubauer} F.~M.,  1980, \mn@doi [\jgr] {10.1029/JA085iA03p01171}, \href
  {https://ui.adsabs.harvard.edu/abs/1980JGR....85.1171N} {85, 1171}

\bibitem[\protect\citeauthoryear{{{\'O} Fionnag{\'a}in} et~al.,}{{{\'O}
  Fionnag{\'a}in} et~al.}{2019a}]{ofionnagain19}
{{\'O} Fionnag{\'a}in} D.,  et~al., 2019a, \mn@doi [\mnras]
  {10.1093/mnras/sty3132}, \href
  {https://ui.adsabs.harvard.edu/abs/2019MNRAS.483..873O} {483, 873}

\bibitem[\protect\citeauthoryear{{{\'O} Fionnag{\'a}in} et~al.,}{{{\'O}
  Fionnag{\'a}in} et~al.}{2019b}]{ofionnagain19-erratum}
{{\'O} Fionnag{\'a}in} D.,  et~al., 2019b, \mn@doi [\mnras]
  {10.1093/mnras/stz1308}, \href
  {https://ui.adsabs.harvard.edu/abs/2019MNRAS.487.3079O} {487, 3079}

\bibitem[\protect\citeauthoryear{{{\'O} Fionnag{\'a}in}, {Vidotto}, {Petit},
  {Neiner}, {Manchester}, {Folsom}  \& {Hallinan}}{{{\'O} Fionnag{\'a}in}
  et~al.}{2021}]{ofionnagain21}
{{\'O} Fionnag{\'a}in} D.,  {Vidotto} A.~A.,  {Petit} P.,  {Neiner} C.,
  {Manchester} W. I.,  {Folsom} C.~P.,   {Hallinan} G.,  2021, \mn@doi [\mnras]
  {10.1093/mnras/staa3468}, \href
  {https://ui.adsabs.harvard.edu/abs/2021MNRAS.500.3438O} {500, 3438}

\bibitem[\protect\citeauthoryear{{P{\'e}rez-Torres} et~al.,}{{P{\'e}rez-Torres}
  et~al.}{2020}]{pereztorres21}
{P{\'e}rez-Torres} M.,  et~al., 2020, arXiv e-prints, \href
  {https://ui.adsabs.harvard.edu/abs/2020arXiv201202116P} {p. arXiv:2012.02116}

\bibitem[\protect\citeauthoryear{{Perger} et~al.,}{{Perger}
  et~al.}{2021}]{perger21}
{Perger} M.,  et~al., 2021, arXiv e-prints, \href
  {https://ui.adsabs.harvard.edu/abs/2021arXiv210310216P} {p. arXiv:2103.10216}

\bibitem[\protect\citeauthoryear{{Plavchan}, {Werner}, {Chen}, {Stapelfeldt},
  {Su}, {Stauffer}  \& {Song}}{{Plavchan} et~al.}{2009}]{plavchan09}
{Plavchan} P.,  {Werner} M.~W.,  {Chen} C.~H.,  {Stapelfeldt} K.~R.,  {Su}
  K.~Y.~L.,  {Stauffer} J.~R.,   {Song} I.,  2009, \mn@doi [\apj]
  {10.1088/0004-637X/698/2/1068}, \href
  {https://ui.adsabs.harvard.edu/abs/2009ApJ...698.1068P} {698, 1068}

\bibitem[\protect\citeauthoryear{{Plavchan} et~al.,}{{Plavchan}
  et~al.}{2020}]{plavchan20}
{Plavchan} P.,  et~al., 2020, \mn@doi [\nat] {10.1038/s41586-020-2400-z}, \href
  {https://ui.adsabs.harvard.edu/abs/2020Natur.582..497P} {582, 497}

\bibitem[\protect\citeauthoryear{{Powell}, {Roe}, {Linde}, {Gombosi}  \& {De
  Zeeuw}}{{Powell} et~al.}{1999}]{powell99}
{Powell} K.~G.,  {Roe} P.~L.,  {Linde} T.~J.,  {Gombosi} T.~I.,   {De Zeeuw}
  D.~L.,  1999, \mn@doi [Journal of Computational Physics]
  {10.1006/jcph.1999.6299}, \href
  {https://ui.adsabs.harvard.edu/abs/1999JCoPh.154..284P} {154, 284}

\bibitem[\protect\citeauthoryear{{Saur}, {Neubauer}, {Connerney}, {Zarka}  \&
  {Kivelson}}{{Saur} et~al.}{2004}]{saur04}
{Saur} J.,  {Neubauer} F.~M.,  {Connerney} J.~E.~P.,  {Zarka} P.,   {Kivelson}
  M.~G.,  2004, {Plasma interaction of Io with its plasma torus}.
pp 537--560

\bibitem[\protect\citeauthoryear{{Saur}, {Grambusch}, {Duling}, {Neubauer}  \&
  {Simon}}{{Saur} et~al.}{2013}]{saur13}
{Saur} J.,  {Grambusch} T.,  {Duling} S.,  {Neubauer} F.~M.,   {Simon} S.,
  2013, \mn@doi [\aap] {10.1051/0004-6361/201118179}, \href
  {https://ui.adsabs.harvard.edu/abs/2013A&A...552A.119S} {552, A119}

\bibitem[\protect\citeauthoryear{{Shulyak} et~al.,}{{Shulyak}
  et~al.}{2019}]{shulyak19}
{Shulyak} D.,  et~al., 2019, \mn@doi [\aap] {10.1051/0004-6361/201935315},
  \href {https://ui.adsabs.harvard.edu/abs/2019A&A...626A..86S} {626, A86}

\bibitem[\protect\citeauthoryear{{Smith}, {Collier Cameron}, {Greaves},
  {Jardine}, {Langston}  \& {Backer}}{{Smith} et~al.}{2009}]{smith09}
{Smith} A.~M.~S.,  {Collier Cameron} A.,  {Greaves} J.,  {Jardine} M.,
  {Langston} G.,   {Backer} D.,  2009, \mn@doi [\mnras]
  {10.1111/j.1365-2966.2009.14510.x}, \href
  {https://ui.adsabs.harvard.edu/abs/2009MNRAS.395..335S} {395, 335}

\bibitem[\protect\citeauthoryear{{Sokolov} et~al.,}{{Sokolov}
  et~al.}{2013}]{sokolov13}
{Sokolov} I.~V.,  et~al., 2013, \mn@doi [\apj] {10.1088/0004-637X/764/1/23},
  \href {https://ui.adsabs.harvard.edu/abs/2013ApJ...764...23S} {764, 23}

\bibitem[\protect\citeauthoryear{{Strugarek}, {Brun}, {Donati}, {Moutou}  \&
  {R{\'e}ville}}{{Strugarek} et~al.}{2019}]{strugarek19}
{Strugarek} A.,  {Brun} A.~S.,  {Donati} J.~F.,  {Moutou} C.,   {R{\'e}ville}
  V.,  2019, \mn@doi [\apj] {10.3847/1538-4357/ab2ed5}, \href
  {https://ui.adsabs.harvard.edu/abs/2019ApJ...881..136S} {881, 136}

\bibitem[\protect\citeauthoryear{{T{\'o}th} et~al.,}{{T{\'o}th}
  et~al.}{2012}]{toth12}
{T{\'o}th} G.,  et~al., 2012, \mn@doi [Journal of Computational Physics]
  {10.1016/j.jcp.2011.02.006}, \href
  {https://ui.adsabs.harvard.edu/abs/2012JCoPh.231..870T} {231, 870}

\bibitem[\protect\citeauthoryear{{Turner} et~al.,}{{Turner}
  et~al.}{2021}]{turner21}
{Turner} J.~D.,  et~al., 2021, \mn@doi [\aap] {10.1051/0004-6361/201937201},
  \href {https://ui.adsabs.harvard.edu/abs/2021A&A...645A..59T} {645, A59}

\bibitem[\protect\citeauthoryear{{Turnpenney}, {Nichols}, {Wynn}  \&
  {Burleigh}}{{Turnpenney} et~al.}{2018}]{turnpenney18}
{Turnpenney} S.,  {Nichols} J.~D.,  {Wynn} G.~A.,   {Burleigh} M.~R.,  2018,
  \mn@doi [\apj] {10.3847/1538-4357/aaa59c}, \href
  {https://ui.adsabs.harvard.edu/abs/2018ApJ...854...72T} {854, 72}

\bibitem[\protect\citeauthoryear{{Vedantham} et~al.,}{{Vedantham}
  et~al.}{2020}]{vedantham20}
{Vedantham} H.~K.,  et~al., 2020, \mn@doi [Nature Astronomy]
  {10.1038/s41550-020-1011-9}, \href
  {https://ui.adsabs.harvard.edu/abs/2020NatAs...4..577V} {4, 577}

\bibitem[\protect\citeauthoryear{{Vidotto} \& {Donati}}{{Vidotto} \&
  {Donati}}{2017}]{vidotto17}
{Vidotto} A.~A.,  {Donati} J.~F.,  2017, \mn@doi [\aap]
  {10.1051/0004-6361/201629700}, \href
  {https://ui.adsabs.harvard.edu/abs/2017A&A...602A..39V} {602, A39}

\bibitem[\protect\citeauthoryear{{Vidotto}, {Fares}, {Jardine}, {Moutou}  \&
  {Donati}}{{Vidotto} et~al.}{2015}]{vidotto15}
{Vidotto} A.~A.,  {Fares} R.,  {Jardine} M.,  {Moutou} C.,   {Donati} J.~F.,
  2015, \mn@doi [\mnras] {10.1093/mnras/stv618}, \href
  {https://ui.adsabs.harvard.edu/abs/2015MNRAS.449.4117V} {449, 4117}

\bibitem[\protect\citeauthoryear{{Vidotto}, {Feeney}  \& {Groh}}{{Vidotto}
  et~al.}{2019}]{vidotto19}
{Vidotto} A.~A.,  {Feeney} N.,   {Groh} J.~H.,  2019, \mn@doi [\mnras]
  {10.1093/mnras/stz1696}, \href
  {https://ui.adsabs.harvard.edu/abs/2019MNRAS.488..633V} {488, 633}

\bibitem[\protect\citeauthoryear{{Wood}, {Linsky}, {M{\"u}ller}  \&
  {Zank}}{{Wood} et~al.}{2001}]{wood01}
{Wood} B.~E.,  {Linsky} J.~L.,  {M{\"u}ller} H.-R.,   {Zank} G.~P.,  2001,
  \mn@doi [\apjl] {10.1086/318888}, \href
  {https://ui.adsabs.harvard.edu/abs/2001ApJ...547L..49W} {547, L49}

\bibitem[\protect\citeauthoryear{{Zaghoo} \& {Collins}}{{Zaghoo} \&
  {Collins}}{2018}]{zaghoo18}
{Zaghoo} M.,  {Collins} G.~W.,  2018, \mn@doi [\apj]
  {10.3847/1538-4357/aac6e8}, \href
  {https://ui.adsabs.harvard.edu/abs/2018ApJ...862...19Z} {862, 19}

\bibitem[\protect\citeauthoryear{{Zarka}}{{Zarka}}{1998}]{zarka98}
{Zarka} P.,  1998, \mn@doi [\jgr] {10.1029/98JE01323}, \href
  {https://ui.adsabs.harvard.edu/abs/1998JGR...10320159Z} {103, 20159}

\bibitem[\protect\citeauthoryear{{Zarka}}{{Zarka}}{2007}]{zarka07}
{Zarka} P.,  2007, \mn@doi [\planss] {10.1016/j.pss.2006.05.045}, \href
  {https://ui.adsabs.harvard.edu/abs/2007P&SS...55..598Z} {55, 598}

\bibitem[\protect\citeauthoryear{{Zarka}, {Treumann}, {Ryabov}  \&
  {Ryabov}}{{Zarka} et~al.}{2001}]{zarka01}
{Zarka} P.,  {Treumann} R.~A.,  {Ryabov} B.~P.,   {Ryabov} V.~B.,  2001,
  \mn@doi [\apss] {10.1023/A:1012221527425}, \href
  {https://ui.adsabs.harvard.edu/abs/2001Ap&SS.277..293Z} {277, 293}

\bibitem[\protect\citeauthoryear{{Zarka}, {Cecconi}  \& {Kurth}}{{Zarka}
  et~al.}{2004}]{zarka04}
{Zarka} P.,  {Cecconi} B.,   {Kurth} W.~S.,  2004, \mn@doi [Journal of
  Geophysical Research (Space Physics)] {10.1029/2003JA010260}, \href
  {https://ui.adsabs.harvard.edu/abs/2004JGRA..109.9S15Z} {109, A09S15}

\bibitem[\protect\citeauthoryear{{de Gasperin}, {Lazio}  \& {Knapp}}{{de
  Gasperin} et~al.}{2020}]{degasperin20}
{de Gasperin} F.,  {Lazio} T.~J.~W.,   {Knapp} M.,  2020, \mn@doi [\aap]
  {10.1051/0004-6361/202038746}, \href
  {https://ui.adsabs.harvard.edu/abs/2020A&A...644A.157D} {644, A157}

\bibitem[\protect\citeauthoryear{{van der Holst}, {Sokolov}, {Meng}, {Jin},
  {Manchester}, {T{\'o}th}  \& {Gombosi}}{{van der Holst}
  et~al.}{2014}]{vanderholst14}
{van der Holst} B.,  {Sokolov} I.~V.,  {Meng} X.,  {Jin} M.,  {Manchester}
  W.~B. I.,  {T{\'o}th} G.,   {Gombosi} T.~I.,  2014, \mn@doi [\apj]
  {10.1088/0004-637X/782/2/81}, \href
  {https://ui.adsabs.harvard.edu/abs/2014ApJ...782...81V} {782, 81}

\makeatother
\end{thebibliography}


\bsp
\label{lastpage}
\end{document}